# Creating mixed-phase states in cadmium sulfide by swift heavy ion irradiation

Aldo Artímez Peña[1,2,*], Nikita Medvedev[1,3]

*1) Institute of Physics, Czech Academy of Sciences, Na Slovance 1999/2, 182 00 Prague 8, Czech Republic*

*2) Higher Institute of Technologies and Applied Sciences, University of Havana, Ave. Salvador Allende 1110, 10 400, La Habana, Cuba*

*3) Institute of Plasma Physics, Czech Academy of Sciences, Za Slovankou 3, 182 00 Prague 8, Czech Republic*

## Abstract

CdS has broad applications in solar cells and radiation detectors. We study its response to irradiation with swift heavy ions and determine its damage thresholds. We apply a model combining the Monte Carlo code TREKIS-3 to simulate the kinetics of the electronic system and the molecular dynamics code LAMMPS to track the atomic reaction to the energy transfer. It is found that ion tracks in CdS differ between its zincblende and wurtzite phases in morphology and damage thresholds. The anisotropic wurtzite phase displays directional damage formation with transient hexagonal track shapes along the (001) plane. In contrast, zincblende CdS exhibits a more cylindrical damage distribution. High pressures near the ion path drive mass transport away from the melted region, forming cavities within the track core in the wurtzite phase. Although significant recrystallization is observed during post-irradiation relaxation, it does not fully restore the original phase: final tracks consist of the amorphous cores due to the low densities in this region and defect-containing crystalline halos. These findings suggest that ion irradiation could be used to create mixed-phase CdS-based materials.

---

[*] Corresponding author: pena@fzu.cz

# I. Introduction

Cadmium sulfide (CdS) belongs to the II-VI semiconductor family[1]. Typical for cadmium chalcogenides, CdS can crystallize in zinc-blende (ZB), wurtzite (WZ), or rock-salt (RS) structures, with the former two being observable under normal conditions[2].

CdS is considered a viable alternative material for space solar cells[3]. It is also used in optoelectronic devices such as light-emitting diodes and radiation detectors[4]. These prospective applications, along with the frequent use of ion beams in optoelectronic materials engineering[5,6], motivate research on the response of CdS-based thin films[7,8], nanostructures[9], and polycrystalline systems to swift heavy ion (SHI) irradiation[10].

An SHI interacts with a solid via electronic (or inelastic) scattering, which excites electrons in the target; nuclear (or elastic) scattering, which transfers kinetic energy to target atoms; and radiative energy losses, emitting photons (such as Bremsstrahlung and Cherenkov radiation)[11]. For ion energies above ~ MeV/a.m.u., the electronic stopping regime is dominant[12]. Heavy ions deposit energy into the electronic system with a stopping power of up to ~35 keV/nm at the Bragg peak[12].

A swift heavy ion triggers cascades of electronic excitation around its trajectory, which lasts for up to ~100 fs. During this time, high-energy electrons ($\delta$-electrons) transport energy away from the core of the track[11]. Quasi-elastic scattering of electrons transfers energy to the atomic lattice. Inelastic electron scattering (impact ionization) generates secondary electrons and holes across various electronic shells[11].

Deep-shell holes created by an SHI or electron impact ionization rapidly decay within a few-fs timescale mainly via the Auger-decay channel emitting an electron. At considered conditions, radiative decay events occur with a lower probability[11,13]. In each decay event, a hole transitions to a higher energy level, ultimately reaching the valence band of the material.

Valence holes, accumulating a significant portion of the deposited energy, behave similarly to the conduction-band electrons (with their own effective mass[11,14]), also delivering energy to the atomic system [11]. In semiconducting materials, such as CdS, where the valence band is broader than the bandgap, inelastic scattering of valence holes can occur, resulting in the excitation of secondary electrons[11,13].

The majority of excited electrons are slow, typically remaining within a distance of ~10 nm from the SHI path within the electronic cooling time. These electrons transiently affect the interatomic potential in that region [11,13]. At the timescale of electronic relaxation in an SHI track, the atoms can be considered static.

By the end of the electron cascade, most of the energy is transferred from excited electrons to the atoms in the material *via* three primary mechanisms: (a) the quasi-elastic scattering of electrons [11,15], (b) the quasi-elastic scattering of valence holes on atoms (coupling to phonons), and (c) atomic acceleration due to nonthermal changes in the interatomic potential induced by the electronic excitation[11,16].

Atomic response to the provided energy may transiently form a disordered region within a few-nanometer radius around an SHI trajectory over the picosecond timescale[11,17]. The momentary increase in the atomic kinetic energy in a highly localized region around an SHI trajectory also increases pressure, emitting shock waves radially outward from the track core which may transiently reduce the material density there[11,18].

Following atomic cooling and relaxation, the damage may remain permanent above a certain energy threshold or may partially or fully recover. In the case of inorganic materials, point defects, dislocations, and accumulated stresses may appear around the track, which may exhibit kinetic activity even after cooling [11].

In this work we study the response of bulk CdS to irradiation with swift heavy ions, identifying the material damage thresholds. We compare the structural modifications in the zincblende and wurtzite phases of CdS. We also evaluate the influence of the direction of SHI impact with respect to the crystallographic orientation of tracks in wurtzite CdS, its anisotropic phase. For these purposes, we apply a multiscale model[11] combining the Monte Carlo (MC) code TREKIS-3[13,15,19] with molecular dynamics (MD) code LAMMPS[20].

## II. Model

We study the effects of SHI irradiation of CdS on the example of Au ion. The hybrid/multiscale model uses the Monte Carlo (MC) code TREKIS-3 and Molecular Dynamic (MD) simulations to describe, respectively, the kinetics of the electronic system including the energy deposition to the lattice, and the atomic response to this energy transfer[11,17].

TREKIS-3 (Time-Resolved Electron Kinetics in SHI-irradiated Solids[19]) code uses asymptotic trajectory event-by-event MC simulations of individual particle propagation to model the effects of irradiation: SHI penetration and ionization of the target; the kinetics of $\delta$- and secondary electrons; the decay of deep shell holes; transport and absorption of the photons produced in the radiative decays of core holes; the valence holes transport and their interaction with target atoms [15]. A comprehensive description of the code is available in Refs.[15,21,22]; below, we briefly recall the methodology employed to model these physical effects.

In the MC, the target is considered as an isotropic and homogeneous atomic arrangement[23,24]. The SHI is simulated as a point-like particle with the effective charge calculated with the Barkas formula[15,25–27]. The free-flight distance of an SHI (as well as of electrons, valence holes, and photons) is sampled using the Poisson distribution [15,26,28]. The mean free path is calculated from the scattering cross-sections obtained with the complex dielectric function (CDF) formalism [16,30]. It allows us to consider collective responses of the electronic system and the lattice (scattering on plasmons and phonons, respectively) [15,29,30]. The CDF parameters for both phases of CdS studied are derived using the single-pole approximation (see Appendix for validation of the cross sections) [31].

Target electrons being ionized occupy either deep atomic energy levels or the valence band according to the density of states (DOS) of the material[32]. The energy transferred to an ionized electron is calculated from the differential scattering cross section [15]. The atomic energy levels (ionization potentials) are taken from Ref. [33].

The propagation and scattering of excited electrons are simulated in the same manner as for an SHI, additionally accounting for the scattering of these electrons on the lattice of the target (quasi-elastic scattering)[34]. Particular realizations of a scattering event—impact ionization of the valence band or atomic shell electrons versus quasi-elastic scattering on the lattice—are chosen according to the partial cross-sections of these processes. No artificial energy cut-off in the tracing of electrons and valence holes is used[15].

Auger and radiative decay times of deep-shell holes are taken from the EPICS2023 database[33]. The shells participating in an Auger decay are chosen randomly proportionally to their electron occupations. The initial energy of the electron equals the difference between the energy released in this hole decay and the electron's ionization potential[15].

The spatial propagation of valence holes is simulated similarly to that of electrons, taking into account the hole's effective mass calculated from the DOS of the valence band within the effective one-band approximation[11,14].

All these processes are traced in TREKIS-3 until 100 fs, after which the density of excited electrons in the center of the track decreases to negligibly small values[11]. The entire MC procedure is averaged over 10,000 iterations to obtain reliable statistics[35].

In CdS, the nonthermal rearrangement of the band structure takes place via lowering of the conduction band levels towards the top of the valence band[36]. Thus, following the previously proposed methodology, the transferred energy due to nonthermal acceleration of atoms can be modeled as a conversion of the potential energy stored in electron-hole pairs into kinetic energy of the atomic lattice[11,31]. The total energy delivered to atoms is, therefore, obtained by adding the potential energy of the electron-hole pairs to the energy transferred to atoms *via* electrons and valence holes elastic scattering, simulated in TREKIS-3[15,17,31,37].

The atomic response to this energy transfer is modeled using classical MD simulation with LAMMPS [20]. The MC-calculated radial energy distributions are used to set the initial conditions for atoms. Initial atomic velocities are generated in cylindrical layers, assuming Gaussian dispersion of the kinetic energies and uniform distribution of momenta.

Stillinger-Weber potential developed by Zhou *et al*[38] is used for both phases of CdS. Melting temperature ($T_{Melt}$) and cohesive energy ($E_c$) values obtained for the zb phase with this potential are 2382 K and -2.76 eV respectively, which are in a reasonable agreement with the experimental data ($T_{Melt}$=2020 K [39], $E_c$=-2.76 eV [38]).

The wurtzite (WZ) supercell used is 47.20 x 40.87 x 20.20 nm$^3$ (1,559,520 atoms) and the zincblende (ZB) supercell is 40.93 x 40.93 x 17.54 nm$^3$ (1,176,000 atoms). Periodic boundary conditions are applied in all directions. The supercell borders in the X- and Y-directions are cooled to 300 K using a Langevin thermostat[40,41] with a characteristic time of 0.1 ps, as the SHI trajectory is set along the Z-axis (the c crystallographic vector in both unit cells). The evolution is tracked up to 250 ps, by which time the maximum temperature in the cell decreases to ~350 K, and therefore, no further structural changes are expected.

For the analysis of different SHI trajectories with respect to the crystallographic orientation, an additional WZ supercell of 45.54 x 20.08 x 45.11 nm$^3$ (1,650,880 atoms) is used. In this case, the ion path is set parallel to the b vector of the wz-CdS unit cell, so the thermostat is applied on X- and Z- borders.

This combined TREKIS+MD approach has been previously validated against experimentally measured damage thresholds and damage kinetics in various irradiated materials[17,42].

Visualization of atomic snapshots and dislocation analysis are performed with OVITO[43].

# III. Results

## 1. Energy transfer to the lattice

Figure 1 shows the radial profiles of the total energy transferred to the lattice, together with the contributions of various channels discussed above. The electron-holes potential energies are comparable in both phases; the differences in the transferred energy arise mainly from the energy transferred to atoms in quasi-elastic scattering. The moderately higher collisional energy calculated for wz-CdS can be attributed to shorter elastic mean free paths of valence band holes (see Appendix, Figure 10).

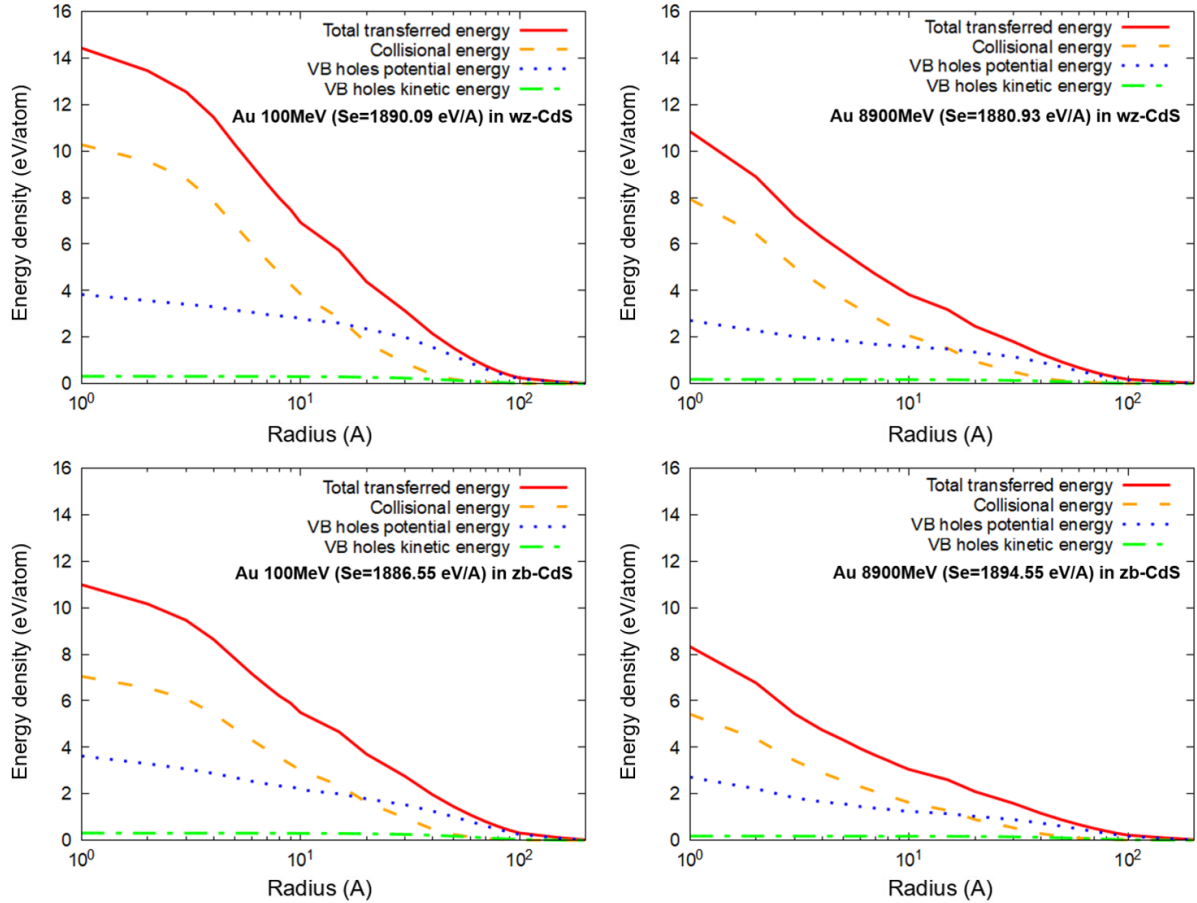

*Figure 1. Radial density of the total energy transferred to the atomic lattice at 100 fs via the three channels.*

Interestingly, within the same energy range (see the electron spectrum in Figure 2), electrons exhibit the opposite behavior: fewer elastic collisions are expected to happen in wz-CdS than in zb-CdS (Appendix, Figure 11), however, this effect is compensated by the higher electron density in the wz phase (Figure 2). In all cases, elastic collisions can be identified as the dominant (but not the only one) mechanism for energy transfer, which is consistent with the thermally induced phase transitions in this material under laser irradiation predicted in Ref. [36].

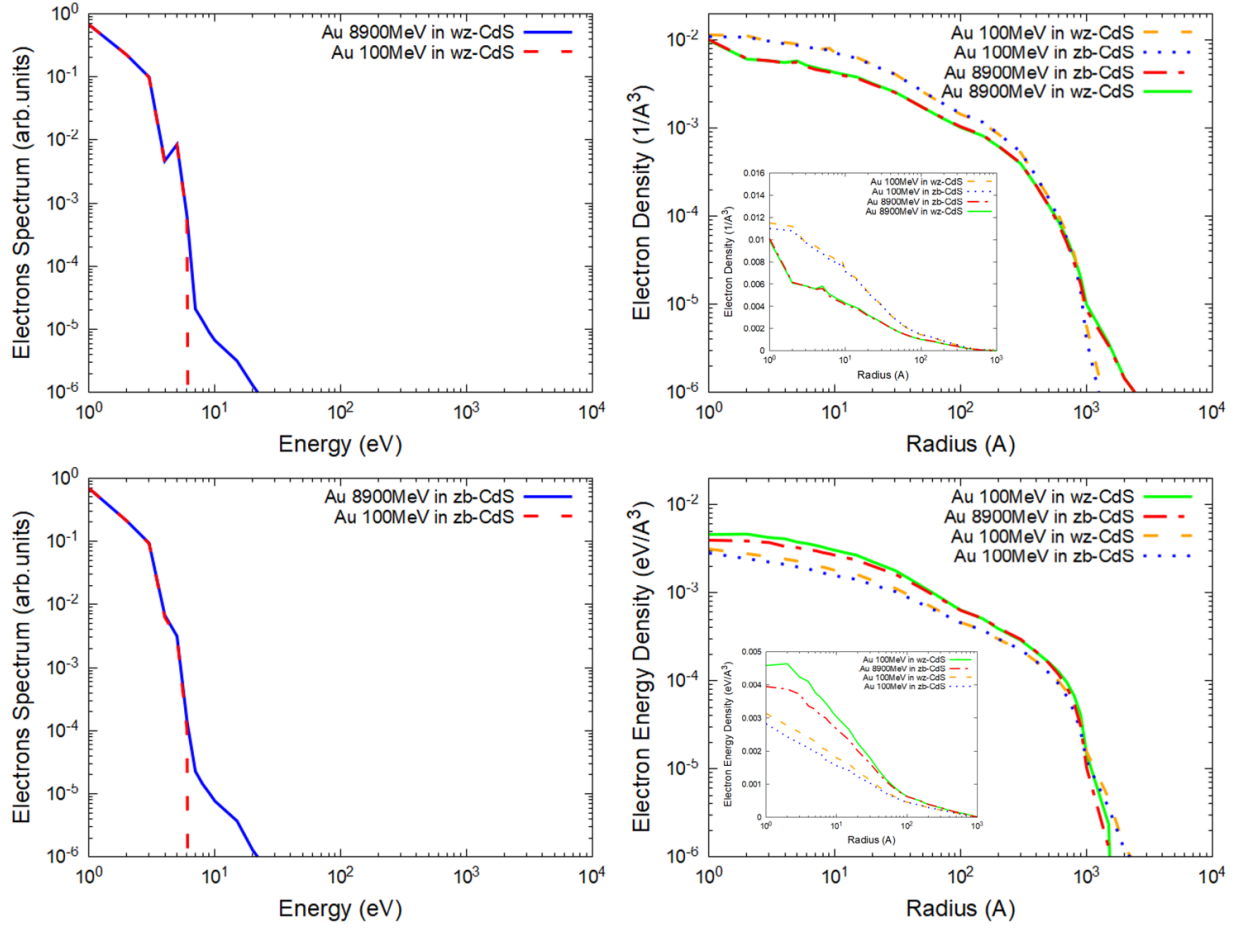

Figure 2. Calculated electron spectrum (a, b), the electron density (c), and the electron energy density (d), in zb-CdS and wz-CdS 100 fs after irradiation with 100 MeV and 8900 MeV Au ion.

As shown in Figure 2, there is a difference in the spectrum of electrons excited after the passage of 100 MeV Au and 8900 MeV Au ions, despite nearly identical ion energy loss (stopping power 18.9 eV/Å, see Appendix): high-velocity Au ion generates more energetic δ-electrons which can travel far from the SHI trajectory, resulting in a lower transferred energy within this region compared to that produced by ions with energies below the Bragg peak with approximately the same energy loss (Figure 1). This is a manifestation of the so-called velocity effect, observed experimentally in various materials [44,45] and discussed in detail, e.g., in [37,46].

## 2. Damage Kinetics

Figure 3 shows the profiles of the atomic temperatures and densities in wz-CdS irradiated with 100 MeV Au ion along the c-axis, along the b-axis, and in zb-CdS. No qualitative differences in the atomic temperature behavior are observed between the two phases; quantitatively, lower temperatures in zb-CdS are due to smaller energy transferred from electrons (cf. Figure 1).

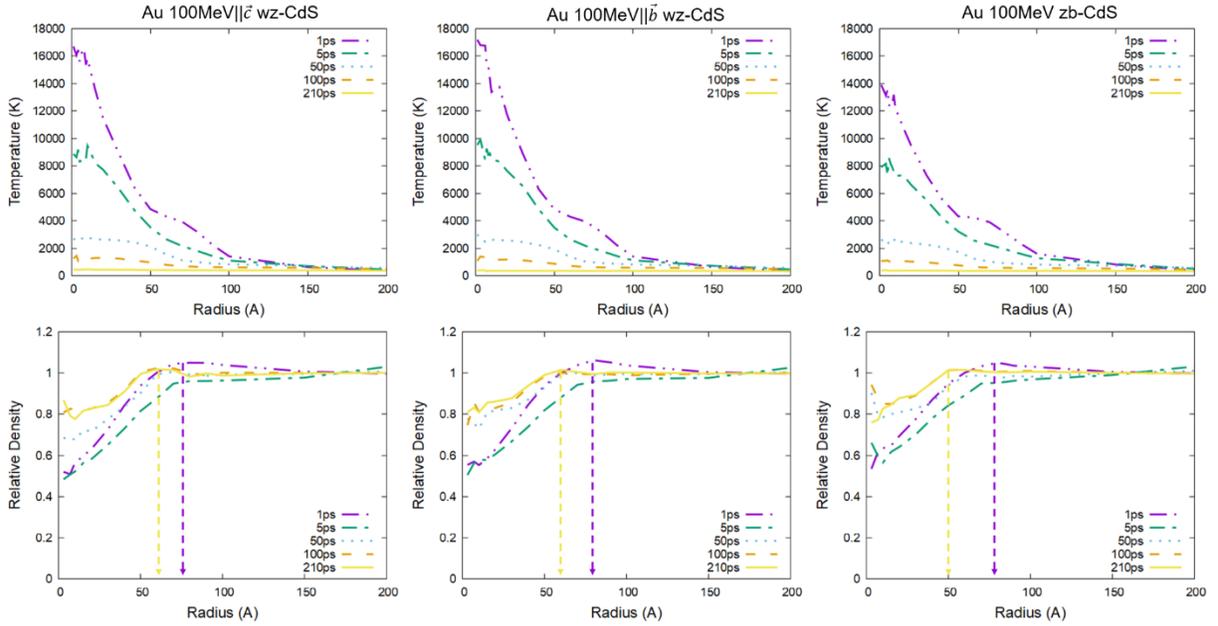

Figure 3. Time-dependent radial distributions of the atomic temperature and density in wz-CdS after irradiation with 100 MeV Au ion along the c-axis (left panel), along the b-axis (middle panel), and in zb-CdS (right panel); dashed arrows indicate the crest of the over-dense front.

The atomic temperature profiles in wz-CdS are nearly identical, independent of the irradiation direction. Note that, similar to other materials [47], the spatial propagation of heat following SHI impact in CdS cannot be described by the parabolic heat diffusion equation due to the presence of transient pressure and density gradients [48].

Figure 3 shows that pressure waves are generated, transporting mass away from the track, resulting in a low-density area in the track's core and slightly over-dense periphery. This density profile persists even after the track cooling; the next section will discuss the corresponding changes in the structure induced.

## 3. Track formation and structural changes

Figure 4 shows the damage evolution in both CdS phases after irradiation with 100 MeV Au ion. The track grows up to ~5 ps after irradiation *via* melting of the surrounding material. This process is very similar in both phases and ion directions; however, structural changes during cooling differ for each case: in wz-CdS irradiated along the c-axis, 1/3<1-210> dislocations tend to loop around the track core, and defects in the track halo form misoriented crystallographic domains (see Appendix, Figure 12).

In contrast, wz-CdS irradiated along the b-axis exhibits 1/3<1-210> dislocations strongly oriented along the a-axis, eventually looping outside the track region. Even though this kind of dislocations is still predominant, 1/3<1-100> and other dislocations coalesce increasing their length in this irradiation direction. In zb-CdS, ½<110> dislocations define the morphology of the periphery of the track.

Track halos also contain point defects besides the misoriented crystallographic domains. Currently, the applied model only accounts for Frenkel pairs generated due to heating, recrystallization and shock waves. However, the contribution of excitons in point defect production can be neglected for CdS, since, given the band gap of the material [49], the energy released by exciton recombination cannot reach the cohesive energy of atoms [38] and therefore is insufficient to displace an atom to interstitial positions[11].

At the end of the simulation, the area of the molten region shrinks by approximately 55%. The final tracks consist of amorphous cores surrounded by defect-containing crystalline peripheries. The recrystallization does not recover the original phase perfectly, but also produces alternative structures. In the case of wz-CdS irradiated along the c-axis, the number of atoms that recrystallize in a zb structure is noticeably lower than in wz-CdS irradiated along the b-axis. Tracks along the b-axis in wz-CdS show a similar degree of recrystallization compared to zb-CdS: after the impact of 100 MeV Au ions, 288 atoms per Å length recrystallized within the zb supercell, while 245 atoms per Å length do so in the wurtzite supercell. However, more atoms in the wz supercell recrystallize into the zb structure than vice versa. These differences in recrystallization depending on the phase and irradiation direction may be important for SHI irradiation in materials engineering since the generation of mixed phases modifies the optoelectronic and photocatalytic properties of CdS-based materials [50,51].

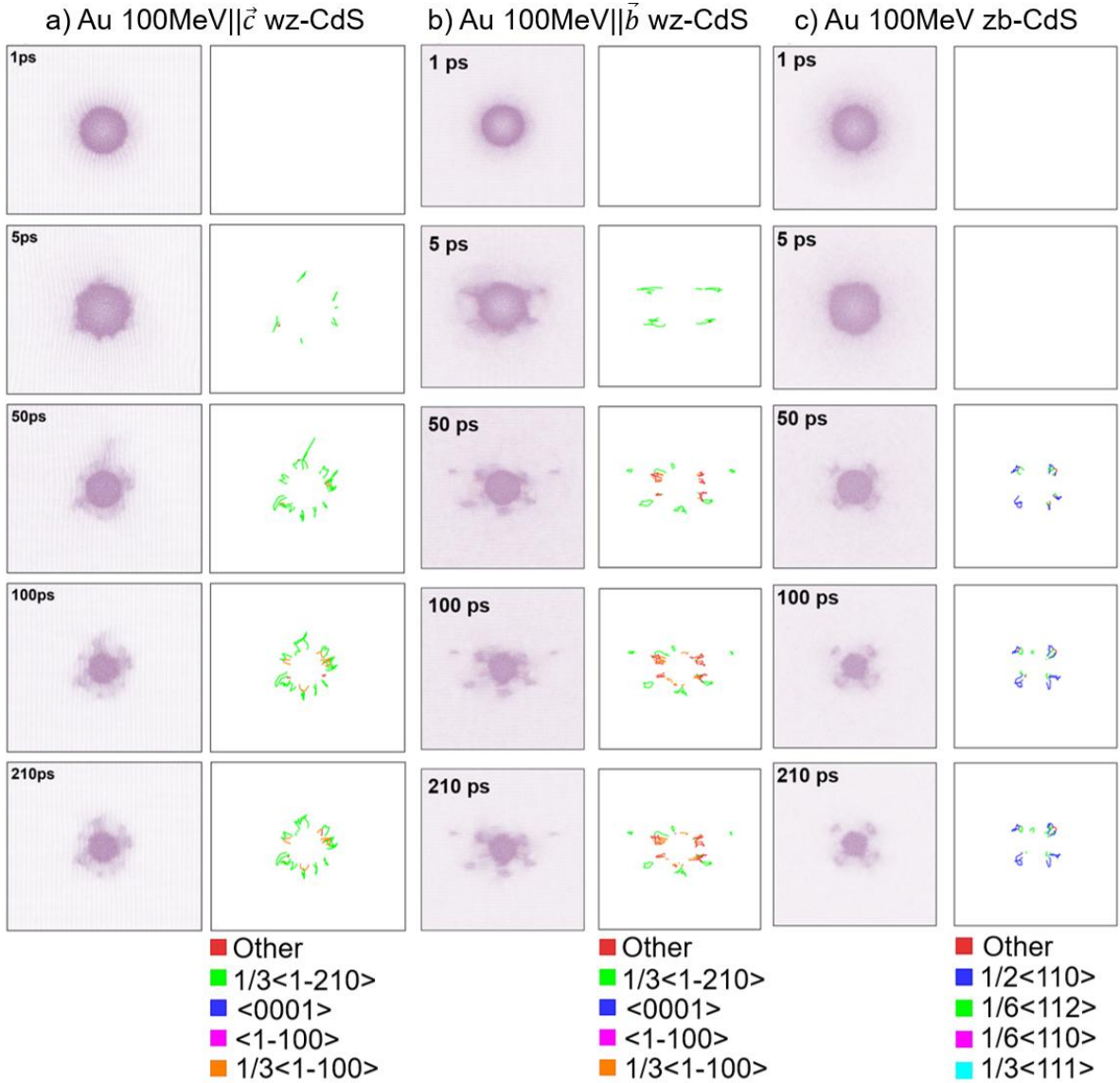

Figure 4. Time evolution of atomic snapshots in tracks (left in each panel) and various types of dislocations (right in each panel, labels at the bottom) generated by 100 MeV Au in wz-CdS along the c-axis (a), wz-CdS along the b-axis (b) and zb-CdS (c).

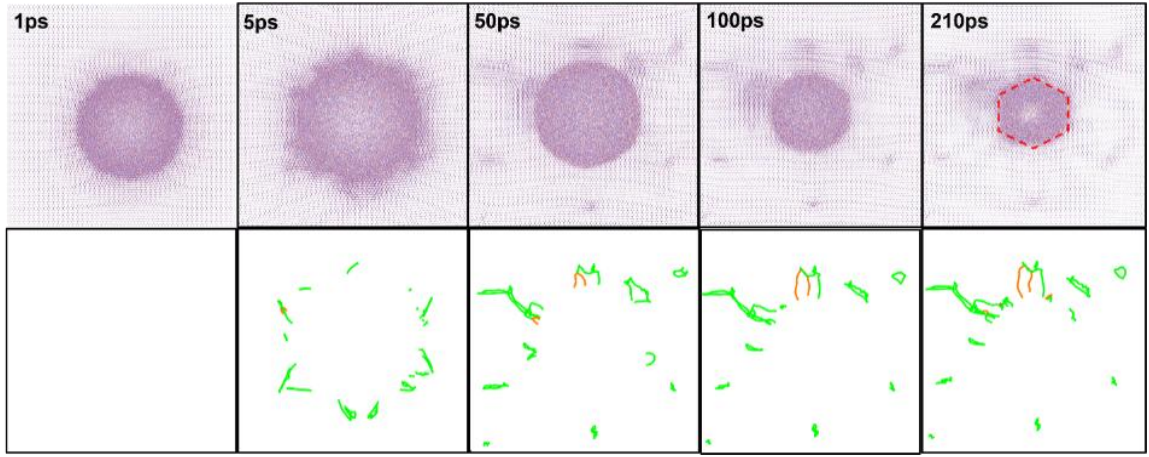

Figure 5. Time evolution of atomic snapshots of tracks (top panels) and dislocations (bottom panels) in wz-CdS after irradiation with 600 MeV Au ion along the c-axis. The dashed hexagon at 210 ps is to guide the eye. 1/3<1-210> dislocations are shown in green, 1/3<1-100> in orange, all other dislocations are shown in red.

At higher ion energies, the track in wz-CdS adopts transient hexagonal geometry along the (001) plane (irradiation with 600 MeV Au ion, shown in Figure 5), similar to those predicted in silicon carbide (SiC) [52]. 1/3<1-210> dislocations are emitted from the hexagonal edges of the melted region. Cavities are formed in the track core only in the case of wz-CdS irradiated along c-axis, but not in the other cases (Figure 6-bottom panel vs. upper panels).

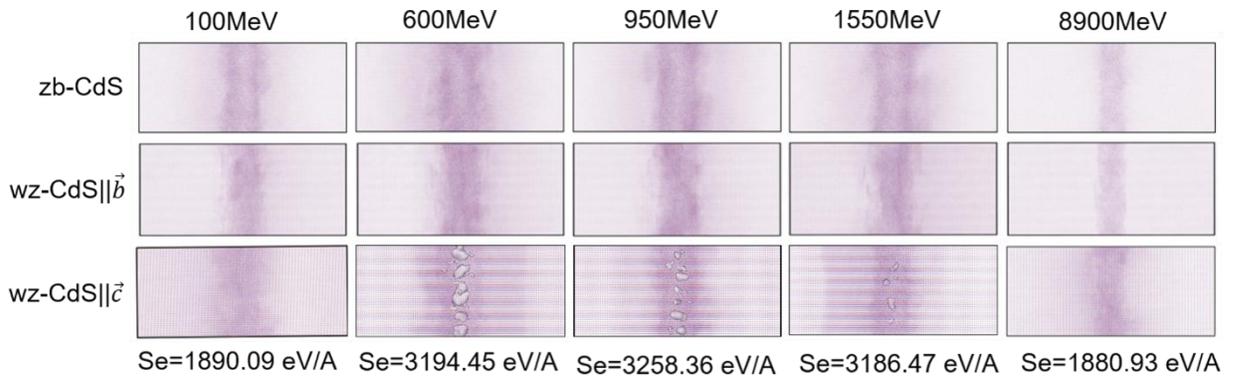

Figure 6. Transversal view of energy-dependent final tracks in zb-CdS (top panel), wz-CdS irradiated along the b-axis (middle panel), and wz-CdS irradiated along the c-axis (bottom panel).

A similar track structure was theoretically and experimentally observed in GaN, a semiconductor with a similar hexagonal structure[53]. It has been suggested that track morphology in GaN is related to density gradients: in low-density regions, recrystallization is not feasible, therefore the track core cools down to an amorphous solid [53]. This also seems to be the case in CdS, since density profiles at different energies correlate with the amorphous regions in the respective tracks as shown in Figure 3.

## 4. Damage threshold

A series of simulations performed for various SHI energies allows us to reconstruct the damage along the ion trajectory and determine the energy range within which the tracks are formed, thereby identifying the ion damage thresholds on the left (slow ions) and right (fast ions) shoulders of the Bragg curve [37]. The estimated damage thresholds are listed in Table 1 and shown in Figure 7 where the track radius curves cross the abscissa.

Table 1. Ion damage thresholds (keV/nm) in wz-CdS and zb-CdS

|  | Slow ion | Fast ion |
|---|---|---|
| SHI$\|\vec{c}$ in wz-CdS | 3.3±1.7 | 13.1±0.3 |
| SHI$\|\vec{b}$ in wz-CdS | 2.8±0.8 | 14.6±0.6 |
| SHI in zb-CdS | 4.1±0.8 | 17.6±0.5 |

Figure 7 shows that the slope in the track radius *versus* energy loss curve in wz-CdS irradiated along the c-axis changes at ~13 keV/nm for slow ions and ~21 keV/nm for fast ions. This indicates the onset of the permanent voids formation in the track core.

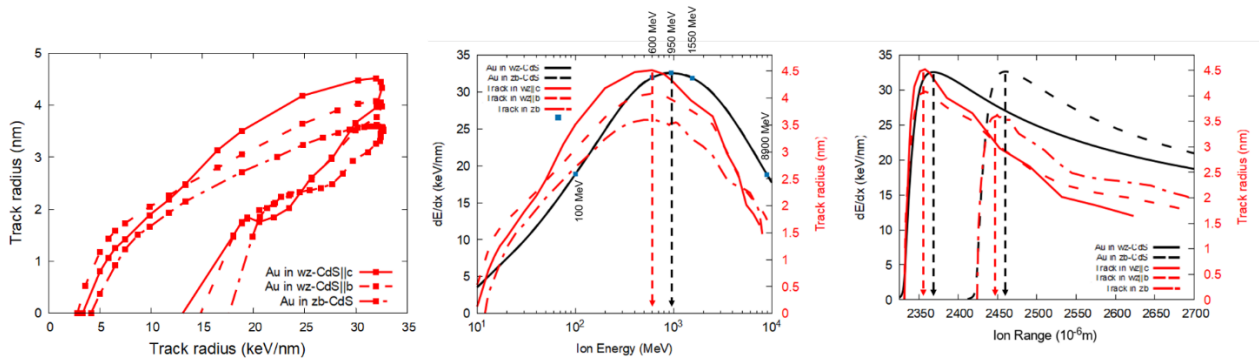

Figure 7. Dependence of the track radius on the energy loss (left panel). Track radius and electronic energy losses vs. ion energy (middle panel) and ion range (right panel).

The higher damage threshold for fast ions (right shoulder of the Bragg curve vs. its left shoulder; see Figure 7) is a direct consequence of the velocity effect discussed above. Fast ions require higher energy loss to reach the transferred energy densities sufficient for track formation [37,46]. The influence of the ion velocity on the final track size can be seen in Figure 8.

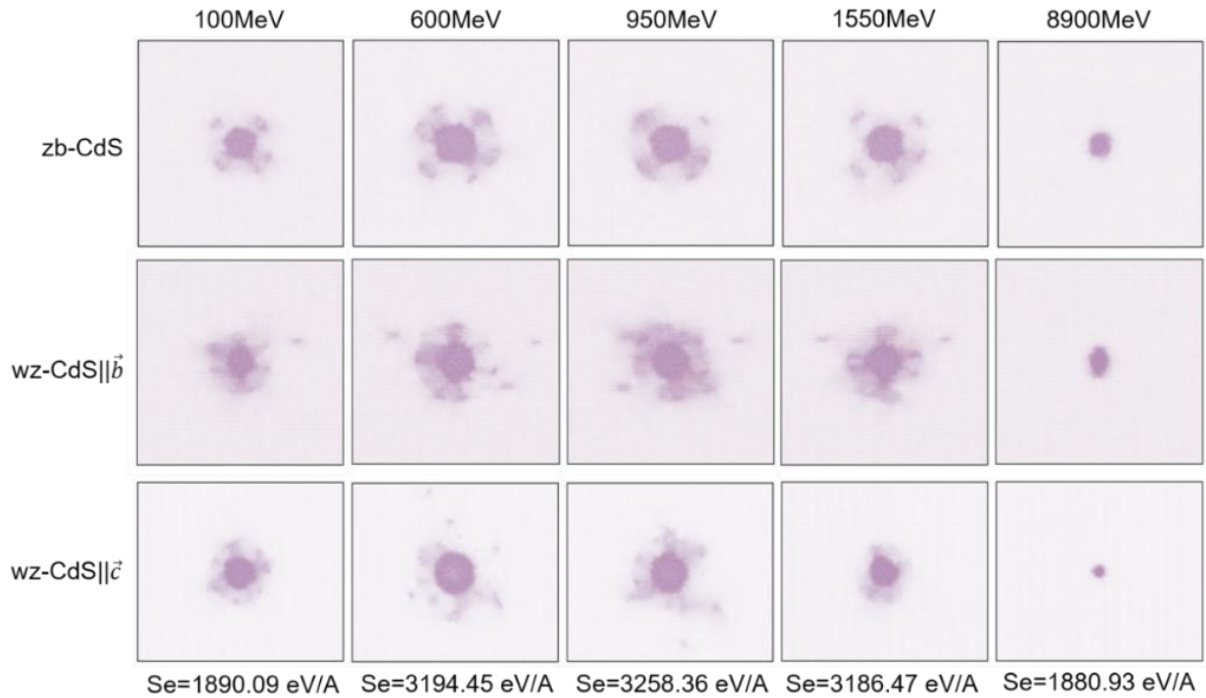

*Figure 8. Energy-dependent final tracks in zb-CdS (top panel), wz-CdS irradiated along the b-axis (middle panel), and wz-CdS irradiated along the c-axis (bottom panel).*

The velocity effect also explains the mismatch between the energy at which the maximal damage is produced (~600 MeV) and the Bragg peak (~950 MeV) in this material (Figure 7). The energy loss defines the total amount of energy deposited into the electronic system and the density of electrons excited by the ion; however, the ion velocity dictates the spectrum of excited electrons and ultimately the profile of energy transferred to the atomic lattice. Therefore, the "optimal" energy for track formation is expected to be in between the Bragg peak and the energy at which maximum energy transfer to the lattice occurs[17,37]. This energy is below the Bragg peak. This also means that the maximal damage is produced deeper along the track than the Bragg peak, see the right panel in Figure 7.

This effect is observed not only in the track size but in other characteristics of the structural damage, for instance, the energy at which the maximal surface area of voids is formed (21,372.5 Å$^2$ at 600 MeV in our simulation box) also differs from the Bragg peak (the voids area at this ion energy is 11,877.8 Å$^2$, Figure 6).

## IV. Conclusions

A comparative analysis of the effects induced by an SHI irradiation in the zinc blende (zb) and wurtzite (wz) phases of CdS was performed using a multiscale approach combining the MC code TREKIS-3 and the MD code LAMMPS. Elastic collisions were identified as the primary mechanism for energy transfer in the track center, consistent with ultrafast XUV/X-ray irradiation simulations of the same material. The higher transferred energy in wz-CdS than in zb phase resulted from shorter elastic mean free paths of valence-band holes. Atomic relaxation produced distinct structural changes in each phase and direction of an SHI impact, leading to amorphous cores with defect-containing crystalline periphery. The recrystallization process does not completely recover the original phase of materials, but significantly reduces the final track size from its transient values. In wz-CdS irradiated along the crystallographic c-axis, the quantity of atoms recrystallizing into a zb-like structure is lower compared to the case of irradiation along the b-axis. This indicates that ion irradiation may be utilized in the creation of mixed-phase CdS-based materials. This process can be tailored by adjusting the irradiation direction. Damage thresholds for slow Au ions were identified as ~2.8 keV/nm (wz-CdS, ion impact along b-axis), 3.3 keV/nm (wz-CdS, impact along c-axis), and 4.1 keV/nm (zb-CdS). For fast Au ions, thresholds were found to be ~14.6 keV/nm (wz-CdS, impact along b-axis), 13.1 keV/nm (wz-CdS, impact along c-axis), and 17.6 keV/nm (zb-CdS). Permanent void formation in wz-CdS irradiated along the c-axis occurs at higher energy loss.

## V. Author contributions (CRediT)

A. Artímez Peña: conceptualization, investigation, formal analysis, visualization, writing -original draft-. N. Medvedev: conceptualization, investigation, formal analysis, visualization, methodology, software, supervision, writing – original draft, writing – review & editing

## VI. Conflicts of interest

There are no conflicts to declare.

## VII. Data and code availability

VIII. The Monte-Carlo code TREKIS-3 used to model SHI irradiation is available from [19].

## IX. Acknowledgments

The authors gratefully acknowledge financial support from the European Commission Horizon MSCA-SE Project MAMBA [HORIZON-MSCA-SE-2022 GAN 101131245]. Computational resources were provided by the e-INFRA CZ project (ID:90254), supported by the Ministry of Education, Youth and Sports of the Czech Republic. NM thanks the financial support from the Czech Ministry of Education, Youth, and Sports (grant nr. LM2023068).

## X. Appendix

The single-pole approximation used for inelastic scattering cross sections of the ion, electrons and valence holes is validated by comparison of the ion stopping power with the SRIM code, and the electron mean free path with the NIST database and other calculations. Both are in good agreement (Figure 9 and Figure 11, respectively), confirming the applicability of CDF formalism and the resulting cross-sections. Figure 10 shows the calculated mean free paths of the valence holes. Figure 12 shows the formation of the misoriented domains, discussed in the main text.

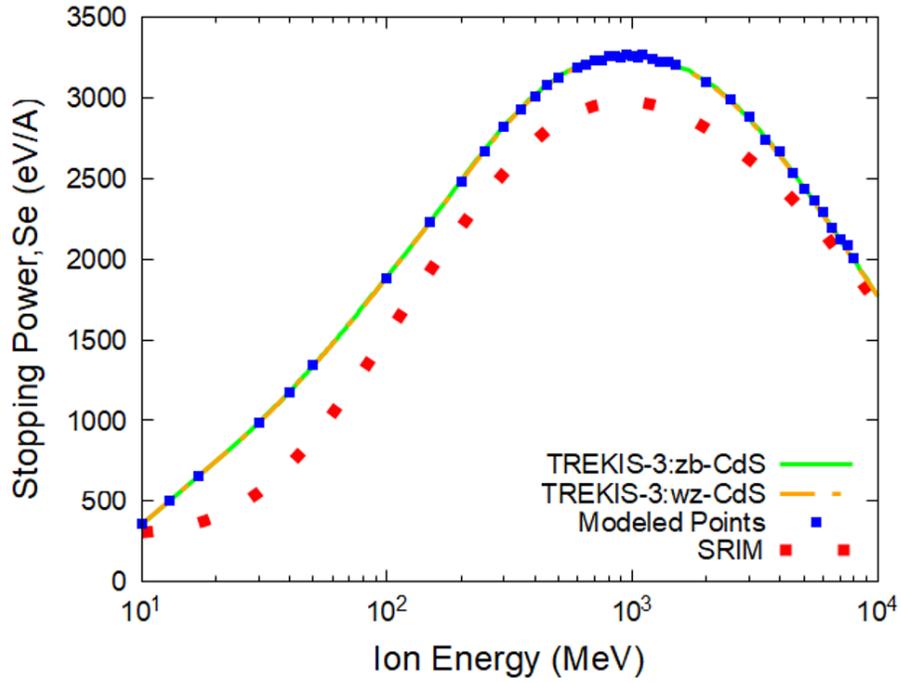

Figure 9. Calculated energy losses of Au ion in zb-CdS and wz-CdS compared with SRIM data [12]. Points mark the energies used for TREKIS+MD modeling.

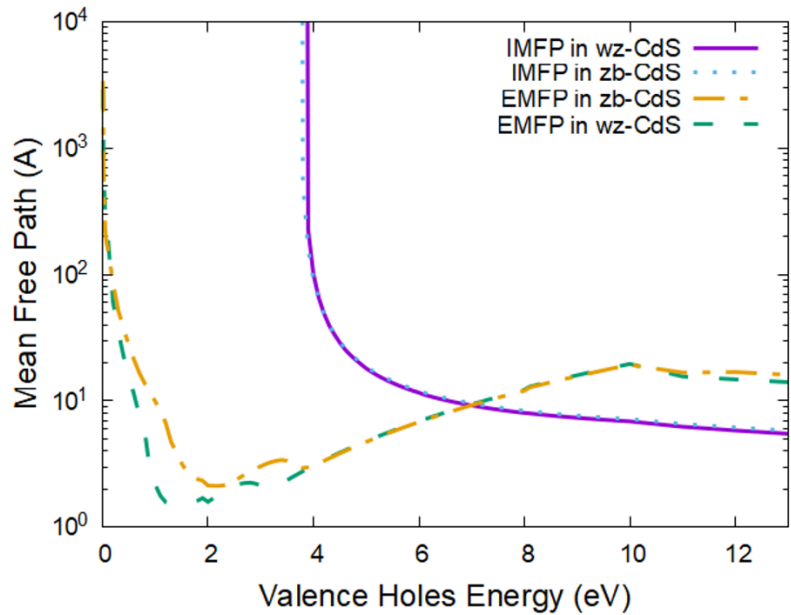

Figure 10. Calculated elastic and inelastic mean free paths of valence band holes in zb-CdS and wz-CdS

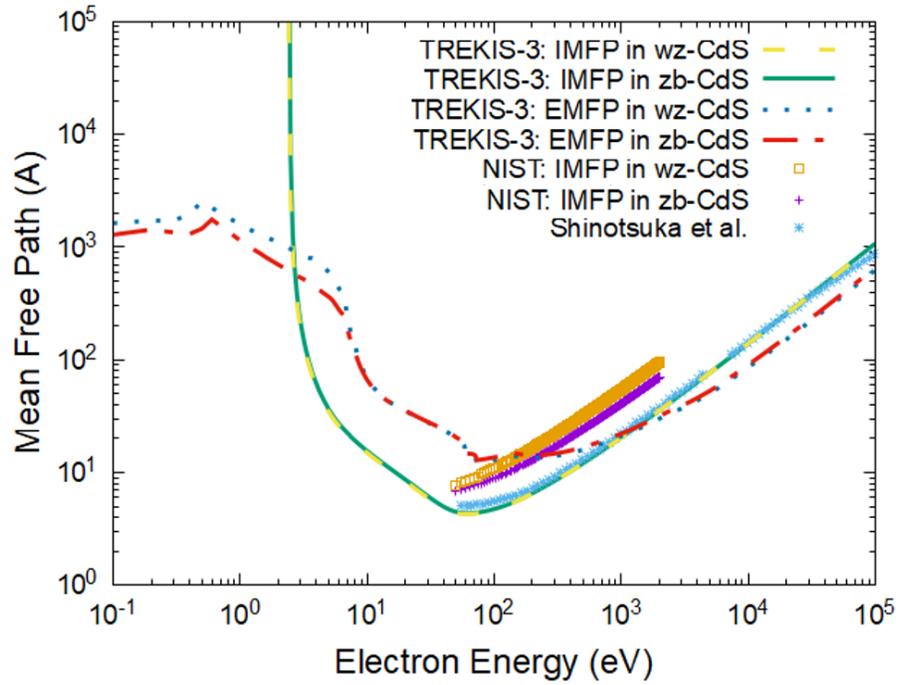

*Figure 11. Calculated elastic mean free path (EMFP) and inelastic ones (IMFP) of electrons in zb-CdS and wz-CdS compared with NIST[54] and Shinotsuka et al. data[55].*

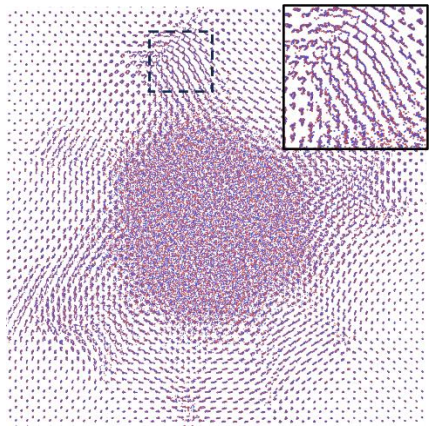

*Figure 12. Final track generated by 100 MeV Au in wz-CdS irradiated along the c-axis. The dashed square highlights a region with misoriented domains (zoomed in right top)*